\documentclass{aa}  

\usepackage{graphicx}
\usepackage{txfonts}
\usepackage{hyperref}
\begin{document}

%
\title{A magnetic reconnection model for the hot explosion with both ultraviolet and H$\alpha$ wing emissions }
\subtitle{}
   \author{Guanchong Cheng
          \inst{1,2,3}
          \and
          Lei Ni\inst{1,3,4,5}
          \and
          Yajie Chen \inst{2}
          \and
          Jun Lin\inst{1,3,4,5}
          }

   \institute{Yunnan Observatories, Chinese Academy of Sciences,Kunming, Yunnan 650216, P. R. China\\
              \email{leini@ynao.ac.cn}
         \and
            Max Planck Institute for Solar System Research, Justus-von-Liebig-Weg 3, 37077, Göttingen, Germany
         \and
             University of Chinese Academy of Sciences, Beijing 100049, P. R. China
         \and
                        Center for Astronomical Mega-Science, Chinese Academy of Sciences, 20A Datun Road, Chaoyang
                        District, Beijing 100012, P. R. China  
         \and
                       Yunnan Key Laboratory of Solar Physics and Space Science, 650216, China
}


 
  \abstract
   {Ellerman bombs (EBs) with significant H$\alpha$ wing emissions and ultraviolet bursts (UV bursts) with strong Si IV emissions are two kinds of small transient brightening events that occur in the low solar atmosphere. The statistical observational results indicate that about 20\% of the UV bursts connect with EBs. While some promising models exist for the formation mechanism of colder EBs in conjunction with UV bursts, the topic remains an area of ongoing research and investigation. }
   {We numerically investigated the magnetic reconnection process between the emerging arch magnetic field and the lower atmospheric background magnetic field. We aim to find out if the hot UV emissions and much colder H$\alpha$ wing emissions can both appear in the same reconnection process and how they are located in the reconnection region. }
   {The open-source code NIRVANA was applied to perform the 2.5D magnetohydrodynamic (MHD) simulation. We developed the related sub-codes to include the more realistic radiative cooling process for the photosphere and chromosphere and the time-dependent ionization degree of hydrogen. The initial background magnetic field is 600 G, and the emerged magnetic field in the solar atmosphere is of the same magnitude, meaning that it results in a low- $\beta$ magnetic reconnection environment. We also used the radiative transfer code RH1.5D to synthesize the Si IV and H$\alpha$ spectral line profiles based on the MHD simulation results.  }
   {Magnetic reconnection between emerged and background magnetic fields creates a thin, curved current sheet, which then leads to the formation of plasmoid instability and the nonuniform density distributions. Initially, the temperature is below 8,000 K. As the current sheet becomes more vertical, denser plasmas are drained by gravity, and hotter plasmas above 20,000 K appear in regions with lower plasma density. The mix of hot tenuous and much cooler dense plasmas in the turbulent reconnection region can appear at about the same height, or even in the same plasmoid. Through the reconnection region, the synthesized Si IV emission intensity can reach above 10$^6$ erg s$^{-1}$ sr$^{-1}$ cm$^{-2}$ \AA$^{-1}$ and the spectral line profile can be wider than 100 km s$^{-1}$, the synthesized H$\alpha$ line profile also show the similar characteristics of a typical EB.  The turbulent current sheet is always in a dense plasma environment with an optical depth larger than 6.5$\times$10$^{-5}$ due to the emerged magnetic field pushing high-density plasmas upward.  }
{Our simulation results indicate that the cold EB and hot UV burst can both appear in the same reconnection process in the low chromosphere, the EB can either appear several minutes earlier than the UV burst, or they can simultaneously appear at the similar altitude in a turbulent reconnection region below the middle chromosphere.}
   \keywords{magnetic reconnection — (magnetohydrodynamics) MHD —solar activity—Sun: 
        heating—Sun: low solar atmosphere—Sun: magnetic flux emergence }
 \maketitle{}
  \titlerunning{}
  \authorrunning{}
%
\section{Introduction}
High-resolution observations across multiple wavelengths have uncovered various small-scale solar activities and their fine structures, both in the quiet Sun and active regions \citep[e.g.,][]{Katsukawa2007,Peter,Huang2017,Toriumi2017,Young,Rouppe,Shen2021,Shen,Joshi}. These small-scale activities are anticipated to provide new insights into the anomalous heating problem of the lower solar atmosphere \citep[e.g.,][]{Parker1988,Peter,Young}. Unlike the upper solar atmosphere, the lower solar atmosphere is cooler and partially ionized, containing both ionized plasma and neutral particles. The interaction between neutral particles and ionized plasma may significantly impact the magnetic reconnection process \citep{Ni2020,Ji2022}, which should be taken into account in the theoretical and numerical studies of solar activities in the low solar atmosphere.\\

Ellerman bombs, discovered in 1917 through spectroscopic observations of H$\alpha$, are characterized by the brightening of the line wings and no significant change in the line center \citep[e.g.,][]{Ellerman,Pariat2004,Pariat2007}. They have a typical size of about one arcsecond and a lifetime ranging from a few tens of seconds to a few minutes \citep[e.g.]{Georgoulis2002,Rutten2013}. Numerous observational studies have shown that EBs also respond strongly in  Ca II h$\&$k lines, Mg II lines, He I lines, and UV continuum \citep[e.g.,][]{Qiu2000,Matsumoto2008,Yang,Vissers,Vissers2015,Tian2016,Hansteen2017,Libbrecht2017}. Most of these spectral lines are low-temperature lines, suggesting that EBs are small, low-temperature solar activities. Several studies based on the inversion of the radiative properties of the H$\alpha$ and Ca II lines using semi-empirical or two-cloud models further indicate that the EBs are usually located in the upper photosphere or the low chromosphere, and the temperature in EBs should be 600-3000 K higher than its surroundings \citep[e.g.,][]{Fang2006,Berlicki2014,Hong2014,Kondrashova2016}. EBs are more likely to occur in active regions near magnetic-field polarity inversion lines, and are accompanied by magnetic flux cancellation, suggesting that their formation is due to magnetic reconnection \citep{Matsumoto2008,Nelson2013,Peter,Reid2016,Tian2016,Shen}.\\

Ultraviolet bursts gained attention after the successful launch of the Interface Region Imaging Spectrograph (IRIS; \cite{De Pontieu2014}) and are characterized by significant brightening in UV images, with a spectral profile displaying strong emission in the Si IV 1394 Å and 1403 Å lines and a weak absorption near the Ni II line \citep[e.g.,][]{Peter,Tian2016,Rouppe van der Voort2017,Tian2018,Young}. These results suggest that UV busts have a high formation temperature (at least on the order of 10$^4$ K), but not a very high formation height. According to most observational studies, UV bursts have a size of approximately one arcsecond, an average duration of a few minutes, and a formation height between the solar temperature minimum region (TMR) and the transition region \citep{Peter,Tian2016,Hou2016}. Magnetic reconnection is also believed to be the primary mechanism responsible for the formation of UV bursts \citep[e.g.,][]{Peter,Tian2016}. In addition to the U-shaped magnetic topology, the fan-spine topology has been discovered in the reconnection region of UV bursts \citep[e.g.,][]{Chitta2017, Chen2019b}.\\

Joint observations between ground-based telescopes and the IRIS satellite indicate that about 20$\%$ of UV bursts connect with EBs \citep[e.g.,][]{Kim2015,Tian2016,Chen2019,Ada2020}. The high-temperature Si IV line emissions and the much cooler H$\alpha$ wing emissions are both observed during the formation process of such UV bursts. Statistical studies of \cite{Chen2019} show that the formation height of the UV bursts can be slightly higher than the corresponding EBs. The UV bursts and the connecting EBs can appear simultaneously, or the UV bursts may appear several minutes after the presence of the connecting EBs \citep{Ada2020}.\\

The formation processes of EBs and UV bursts have been numerically studied based on the MHD simulations. In the early simulations of \cite{Isobe2007} and \cite{Archontis2009}, the upwelling magnetic field due to the Parker instability forms a sea-snake-like magnetic field, and magnetic fields on opposite sides of a U-loop structure reconnect to form EBs, which is consistent with the scenarios from observations in \cite{Pariat2004}. In \cite{Danilovic2017}, they synthesized the H$\alpha$ spectral line profiles and the H$\alpha$ wing images of the simulated EBs based on the more realistic radiation MHD simuations, their results are in good agreement with the observational features of EBs.  For the first time, \cite{Ni2015} showed that the plasma temperature  in the chromosphere can be heated above 20,000 K during a low- $\beta$ magnetic reconnection process with strong magnetic fields. The further single-fluid and multi-fluid MHD simulations prove that the high-temperature UV bursts can indeed appear in the low- $\beta$ magnetic reconnection process in the chromosphere \citep{Ni2016,Ni2018,Peter2019}.\\

Recently, a few MHD simulation works have focused on explaining the formation mechanisms of the UV bursts connecting with EBs. The 3D Radiation MHD simulations in \cite{Hansteen2019} show that the hot upper part and the cool lower part of a nearly vertical current sheet in the low solar atmosphere correspond to the UV burst and the EB, respectively. Such a current sheet is located in a large, emerging, magnetic-field bubble, being carried with cold gas from the photosphere, and the cold dense plasma surrounding the UV burst can explain the formation of Ni II absorption lines and the absence of intense emission from AIA 304 A channel. This scenario provides a reasonable explanation for the formation of the coexisting UV bursts and EBs. Recent high-resolution simulations of magnetic reconnection between the strong emerging and the background magnetic fields have shown that the multiple plasmoids and turbulent structures appear after the current sheet is emerged to the low solar chromosphere \citep{Ni2021}.  The hot plasmas with lower density and cool plasmas with higher density can both appear at the same altitude in the low chromosphere, and the cool plasmas inside the current sheet extends downward to a lower height, as reported by \cite{Ni2021}. This magnetic reconnection model provides another possible explanation for UV bursts associated with EBs; additionally, the occurrence of hot UV bursts in the low solar chromosphere can naturally account for the observed absorption in Ni II lines and the lack of emission from the AIA 304 A channel. The simulations in \cite{Ni2021} investigated the fine structures within the current sheet well, but the time-dependent ionization degree is not included, and the radiative cooling model is relatively simple. \\

In this work, we explored the formation mechanism of coexisting UV bursts and EBs based on more realistic MHD simulations. Compared with previous work by \cite{Ni2021}, the time-dependent ionization degree of hydrogen and more realistic radiative cooling models for the photosphere and chromosphere are included. Furthermore, we synthesized the Si IV and H$\alpha$ spectral line profiles based on the MHD simulations using the radiative transfer code RH1.5D \citep{Pereira2015}. The remainder of this paper is structured as follows. Section 2 describes the numerical model and the initial and boundary conditions. Section 3 presents the numerical results. A summary and discussion are given in Section 4.

\section{Numerical setup}

To simulate activity phenomena related to EB and UV bursts, we employed a modified version of NIRVANA3.8. Compared to previous work, our approach includes two key improvements. Firstly, we account for the variation in the ionization degree with plasma parameters, thereby providing a more accurate representation of the impact of ionization degree, magnetic dissipation coefficient, and ambipolar diffusion coefficient on simulation results. Secondly, we enhance the radiation cooling model to better capture the influence of radiation cooling on active phenomena in the chromospheric environment. The MHD equations solved in this study are formally identical to those used in the previous work of  \cite{Ni2021}. \\

\subsection{Ionization degrees and diffusivities}

The incorporation of the time-dependent ionization degree is one of the most significant improvments to this simulation. When the plasma parameters are known, the ionization degree of the plasma in conditions of local thermodynamic equilibrium (LTE) can be determined from the Saha equation and the Boltzmann equation; however, the chromospheric environment meets neither the LTE assumption nor the fully ionized coronal equilibrium. To calculate the degree of ionization of the chromospheric environment, the non-LTE assumption must be taken into account. Unfortunately, the exact calculations necessitate computing the distributions of each energy level state and solving them iteratively, which is very computationally expensive. Therefore, we adopted a compromise approach. In the region close to the photosphere, the hydrogen ionization degree $Y_{i}$ is calculated using the Saha equation; in the lower density chromospheric plasma environment, we obtain the temperature-dependent ionization degrees using the table supplied by \cite{Carlsson2012}. 

The values of ionization degree at each half time step will be updated according to the local plasma parameters, and in our simulations the magnetic dissipation coefficient and the ambipolar diffusion coefficient are also related to the ionization settings. The magnetic dissipation coefficient was slightly improved in our simulations, and the physical magnetic diffusion coefficient used is derived from \cite{Collados2012} and expressed as follows:
\begin{eqnarray}
&&\eta=\eta_{\mathrm{ei}}+\eta_{\mathrm{en}}=\frac{m_{\mathrm{e}} v_{\mathrm{ei}}}{e_{\mathrm{c}}^2 n_{\mathrm{e}} \mu_0}+\frac{m_{\mathrm{e}} v_{\mathrm{en}}}{e_{\mathrm{c}}^2 n_{\mathrm{e}} \mu_0}
,\end{eqnarray}
where $m_{\mathrm{e}}$ is the mass of the electron, $e_{\mathrm{c}}$ is the electron charge, $n_{\mathrm{e}}$ is the electron density, and $v_{\mathrm{ei}}$, and $v_{\mathrm{en}}$ are frequencies of electron-ion and electron-neutral collisions, respectively. Finally, using a similar simplification to \citet{Ni2022}, we obtain the following formulas for $\eta_{\mathrm{ei}}$ and $\eta_{\mathrm{en}}$:
\begin{eqnarray}
&&\eta_{\mathrm{ei}} \simeq 1.0246 \times 10^8 \Lambda T^{-1.5} ,\\
&&\eta_{\mathrm{en}} \simeq 0.0351 \sqrt{T} \frac{1-Y_{\mathrm{i}}}{Y_{\mathrm{i}}},\\
&&\Lambda=23.4-1.15 \log _{10} (\frac{\rho}{m_\mathrm{i}} Y_{\mathrm{i}})+3.45 \log _{10} T
.\end{eqnarray}\\
We took the same approach as in the previous work \citep{Ni2021} to obtain the expression for the ambipolar diffusion coefficient:
\begin{eqnarray}
&&\eta_{\mathrm{AD}}=1.65 \times 10^{-11}\left(\frac{1}{Y_i}-1\right) \frac{1}{\rho^2 \sqrt{T}}
.\end{eqnarray}\\

\subsection{Radiation cooling models}
        
The energy-exchange processes between the solar atmosphere and the radiation field, also called radiative transfer processes, significantly influence the behavior and characteristics of the plasma in the atmosphere, especially in the photospheric and chromospheric layers. The radiative transfer processes must be better handled in MHD simulations to realistically simulate the solar activities due to the fact that our simulated region consists mainly of the solar photosphere and chromosphere. However, applying realistic radiative transfer models to MHD simulations is difficult and requires huge computational resources, so we considered separate radiative cooling models to represent the cooling in the photosphere and chromosphere layers, bridging the two by density thresholds to make a relatively complete model. As the simulated region in this study only encompasses the photosphere and chromosphere regions, optically thin radiation models suitable for the coronal region were not incorporated. \\

The first model, which is applied to the region in the photosphere in this simulation, is an approximation proposed by \citet{Abbett2012} that can be substituted for the radiative cooling process in regions with a greater optical depth. The formula for this method is given by 
        \begin{eqnarray}
        &&Q_{r a d 1} \approx-2 \kappa^{B} \rho \sigma T^{4} E_{2}\left(\tau^{B}\right)
        ,\end{eqnarray}\\
        where $\kappa^{B}$ is the Planck-averaged opacity that dependents on the plasma density and temperature, $\tau^{B}$ the optical depth computed from this opacity, $\sigma$ $=5.670\times 10^{-8} \mathrm{Js}^{-1} \mathrm{~m}^{-2} \mathrm{~K}^{-4}$ the Stefan-Boltzmann constant, and $E_{2}$ the exponential integral.\\
        
        
The second model, also an approximation proposed by \citet{Carlsson2012}, is used in the chromosphere region in this simulation, which focuses on the effect of radiative cooling in several important spectral lines of the chromosphere, such as H I Ca \uppercase\expandafter{\romannumeral2}, and Mg \uppercase\expandafter{\romannumeral2}. This model can be described by the equation below:\\
        \begin{eqnarray}
        &&Q_{r a d 2}=-\sum_{X=H, M g, C a} L_{X m}(T) E_{X m}(\tau) \frac{N_{X m}}{N_{X}}(T) A_{X} \frac{N_{H}}{\rho} n_{e} \rho
        .\end{eqnarray}\\
        Here, $L_{X_{m}}$ is the optically thin radiative loss function per electron and per particle of element $X$ in ionization stage $m, E_{X_{m}}(\tau)$ is the photon escape probability as a function of the depth parameter $\tau, \frac{N_{X_{m}}}{N_{X}}$ is the fraction of element $X$ in ionisation stage $m$, and $A_{X}$ is the abundance of element X. In this simulation, $X$ comprises $H$, $Ca,$ and $Mg$; with the exception of $A_{x}$, most of the parameters may be determined by looking up the table offered by \cite{Carlsson2012}.  According to the model developed by \citet{Avrett2008}, $A_{H}$ = 1, $A_{Mg}$ = $3.388 \times10^{-5}$ and $A_{Ca}$ = $2.042\times10^{-6}$. $\frac{N_{H}}{\rho}=4.407\times10^{23}$ g$^{-1} $is the number of hydrogen particles per gram of chromospheric material. \\
        
        Therefore, the radiative cooling term in the energy equation is given as\\
        \begin{eqnarray}
        \begin{aligned}
         &&L_{rad} &= \begin{cases}Q_{r a d 1}, & N_{H}> N_{H0} \\
        Q_{r a d 2}, & N_{H}\leqslant N_{H0} \end{cases} \\
        \end{aligned}
        .\end{eqnarray}
        Here, $N_{H}$ is the number density of the element hydrogen, and $N_{H0}$ =$1\times 10^{21}$ m$^{-3}$.We also specify a heating term H to ensure that the simulation begins in equilibrium. At t = 0s, the heating term is equal to the cooling term, as $H_{0}$ = $-L_{rad0}$. Then, the heating term drops  exponentially with time to smaller and smaller values. Fig.1 shows the initial distribution of the radiative cooling term at different heights. 
           \begin{figure}
   \centering
   \includegraphics[width=8cm]{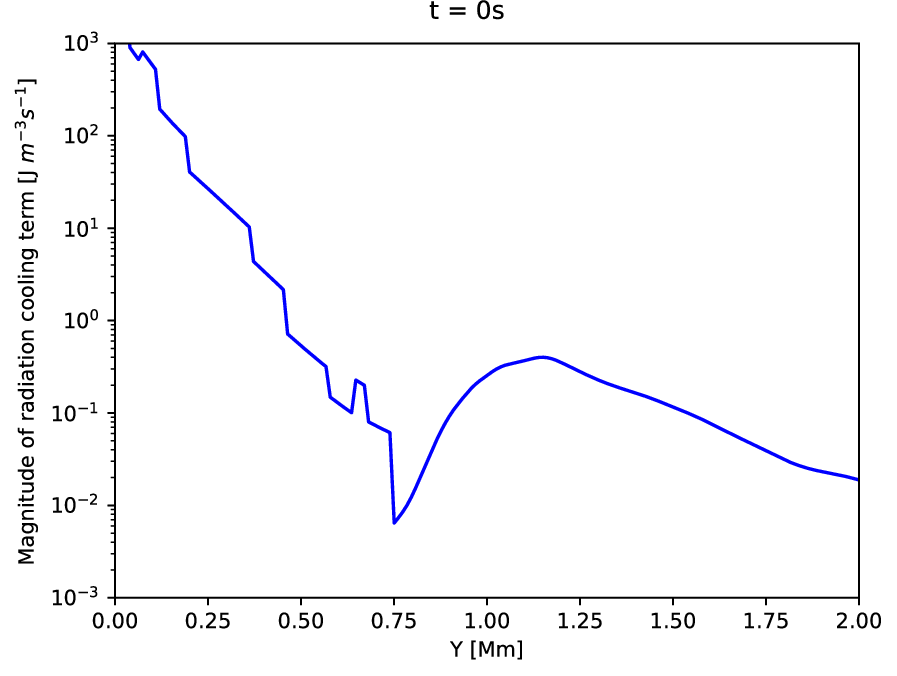}
      \caption{Initial distributions of the radiative cooling term along the vertical direction.}
         \label{FigVibStab}
         \end{figure}
        
           \begin{figure}
   \centering
   \includegraphics[width=8cm]{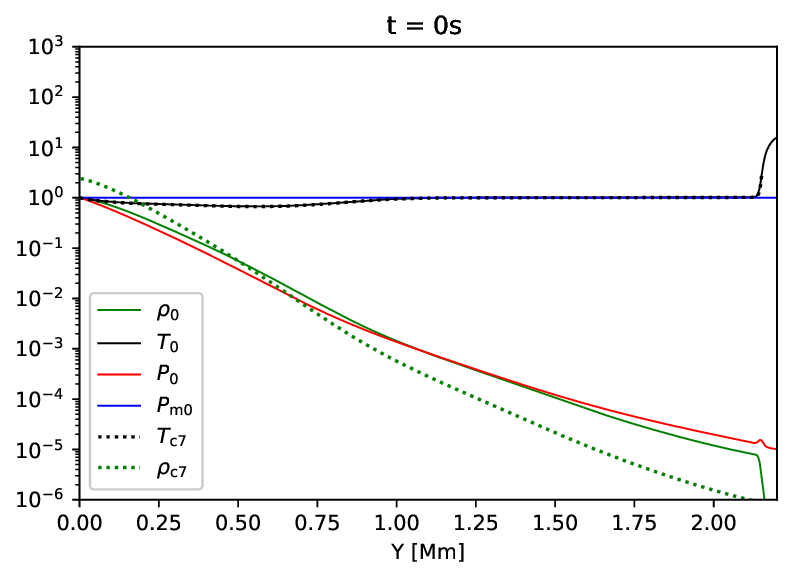}
      \caption{Initial distributions of the temperature, plasma density, thermal pressure, and magnetic pressure along the vertical direction. The variables are normalized by using the reference values at $\mathrm{y}=0 \mathrm{~km}$. These reference values are $\varrho_{\text {ref }}=10^{-3.72}$ $\mathrm{kg} \mathrm{m}^{-3}, T_{\text {ref }}=10^{3.84} \mathrm{~K},P_{\text {ref }}=10^{4.05} \mathrm{~Pa}, P m_{\text {ref }}=$ $10^{3.6} \mathrm{~Pa}$. The dotted black line and the dotted green line represent the distributions of the temperature and plasma density in the C7 atmosphere model, respectively.}
         \label{FigVibStab}
         \end{figure}
\subsection{The initial setups}
        
This simulation domain extends from -8 Mm to 8 Mm in the x direction and from 0 to 2.2 Mm in the y direction that is perpendicular to the solar surface. The initial grids are 384$\times$384, and the maximum adaptive mesh refinement level is 5. At each level of refinement, the resolution is increased by a factor of 2, so the smallest grid size is 651 m in the x direction and 179 m in the y direction.\\
        
The initial values of the temperature and ionization degree in the y direction are obtained from the C7 solar atmosphere model \citep{Avrett2008},  and the initial density distribution in the y direction is calculated according to the hydrostatic equilibrium equation, $\nabla \mathrm{p} - \rho \mathbf{g}=0$. Fig. 2 shows the initial distributions of these variables in the y direction.\\

To simulate the reconnection process between the background magnetic field and the emergent arching magnetic field, we created a uniform background magnetic field that covers the full simulation region and modify the bottom boundary condition to achieve a progressively increasing arching magnetic field. The background field at the initial moment is set as follows:
        \begin{eqnarray}
        &&B_{x 0}=-0.85 b_{0} \nonumber\\
        &&B_{y 0}=-0.35 b_{0} \\
       , &&B_{z 0}=c_{0} b_{0}\nonumber
        \end{eqnarray}
        where $B_{x 0}$,$B_{y 0}$, and $B_{z 0}$ are the initial magnetic field components in the spatial dimensions; $c_{0}$ and $b_{0}$ are constants. We chose $c_{0}=1$ and $b_{0}$=600 G for this simulation.\\
        
        The following equations, which were similar for \citet{Ni2021}, describe the bottom boundary conditions of magnetic field and are utilized to realize an emergent arched field:
        
        \begin{equation}
        \begin{aligned}
        B_{x b} &= \begin{cases}-0.85 b_{0}+b_{1} \frac{\left(y-y_{0}\right) L_{0}^{1.6} t}{ \left[x^{2}+\left(\left(y-y_{0}\right)^{2}\right]^{1.3} t_{0}\right.}, & t \leqslant t_{0} \\
        -0.85 b_{0}+b_{1} \frac{\left(y-y_{0}\right) L_{0}^{1.6}}{\left[x^{2}+\left(\left(y-y_{0}\right)^{2}\right]^{1.3}\right.}, & t>t_{0}\end{cases} \\
        B_{y b} &= \begin{cases}-0.35 b_{0}-b_{1} \frac{x L_{0}^{1.6} t}{\left[x^{2}+\left(\left(y-y_{0}\right)^{2}\right]^{1.3} t_{0}\right.}, & t \leqslant t_{0} \\
        -0.35 b_{0}-b_{1} \frac{x L_{0}^{1.6}}{\left[x^{2}+\left(\left(y-y_{0}\right)^{2}\right]^{1.3}\right.}, & t>t_{0}\end{cases} \\
        B_{z b} &=b_{0}
        \end{aligned}
        ,\end{equation}
where $t_{0}=100 \mathrm{~s}$, $b_{1}$=600 G, $y_{0}$=10$^6$ m, and $L_{0}$=10$^6$ m. The implementation of such a boundary condition for the magnetic fields at the bottom leads to the emergence of arched magnetic fields. As the simulation progresses beyond 100 s, an arch-shaped magnetic field gradually emerges into the photosphere from the lower boundary. To better understand this boundary condition's functionality, we recommend viewing the movie provided in the supplemental file. The average emergence velocity of the arched magnetic field is 10 km$\cdot$$s^{-1}$, a value that is consistent with magnetic field emergence velocities found in previous research works \citep[e.g.,][]{cheung2014}. This approach also aids in conserving computational resources. For all the variables, we used exactly the same boundary conditions as those in the previous work \citep{Ni2021}.\\

%
%

\begin{figure*}
        \centering
        \includegraphics[width=20cm]{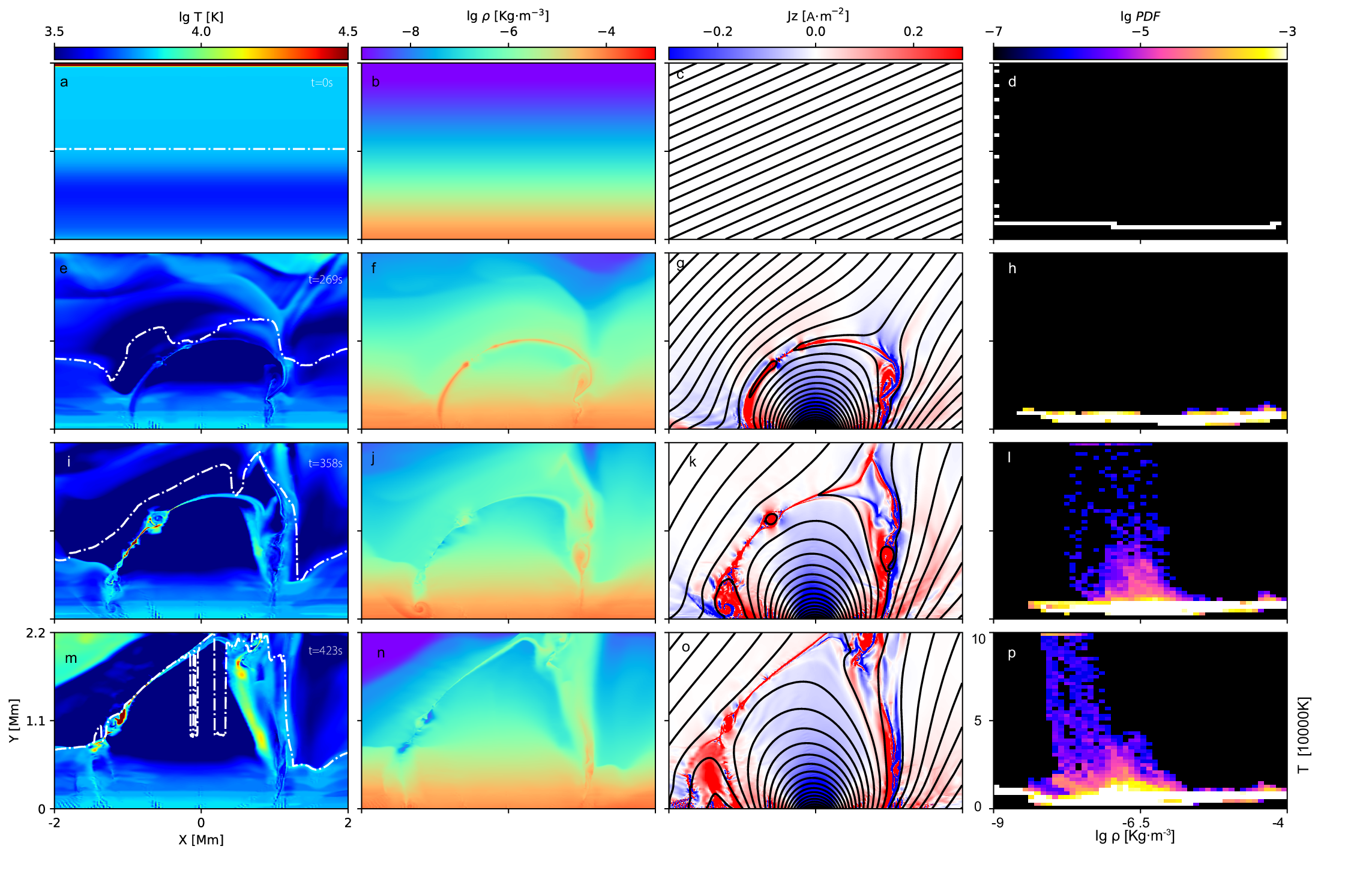}
        \caption{General view of simulation results at different moments. Columns 1, 2, and 3 display the temperature, density, and current density ($J_{z}$) distributions for four different times, respectively. Column 4 represents the PDF of temperature and density at various times. The white dashed lines in Column 1 correspond to the mean optical depth, that isn equal to $6.5\times10^{-5}$, and the black solid lines in Column 3 represent the magnetic field lines.}
        \label{FigGam}%
\end{figure*}

\begin{figure*}
        \centering
        \includegraphics[width=18cm]{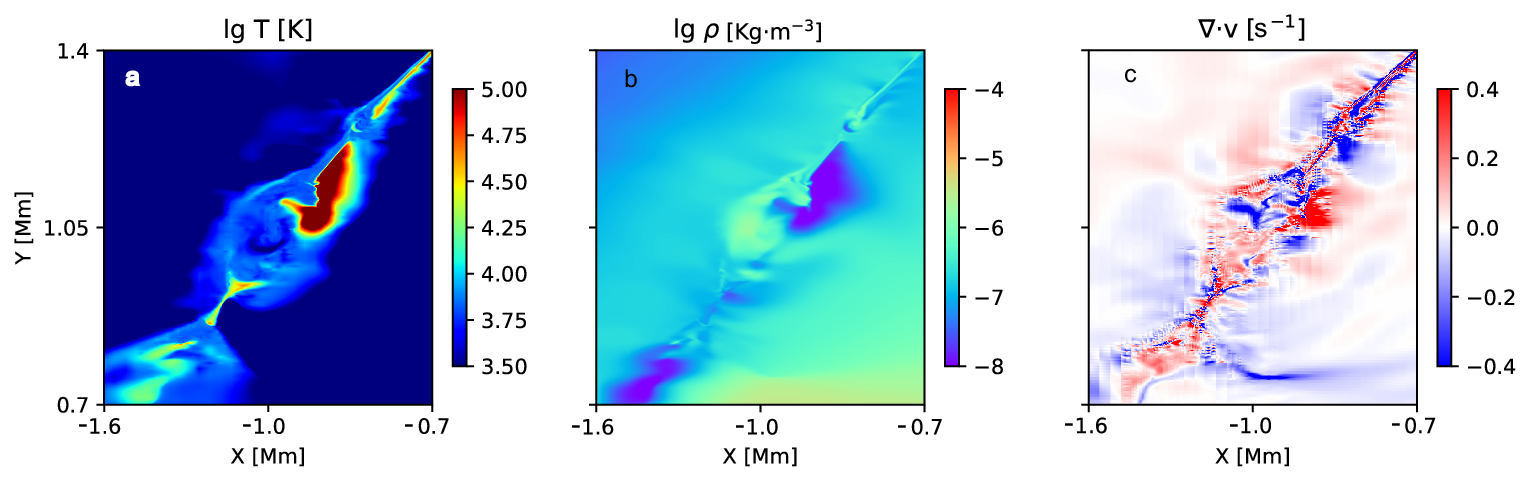}
        \caption{Zoomed-in view of plasmoid region at t = 423s. Panel a shows the temperature distribution; panel b is the density distribution; panel c shows the distribution of the velocity divergence.}
        \label{FigGam}%
\end{figure*}

\begin{figure*}
        \centering
        \includegraphics[width=18cm]{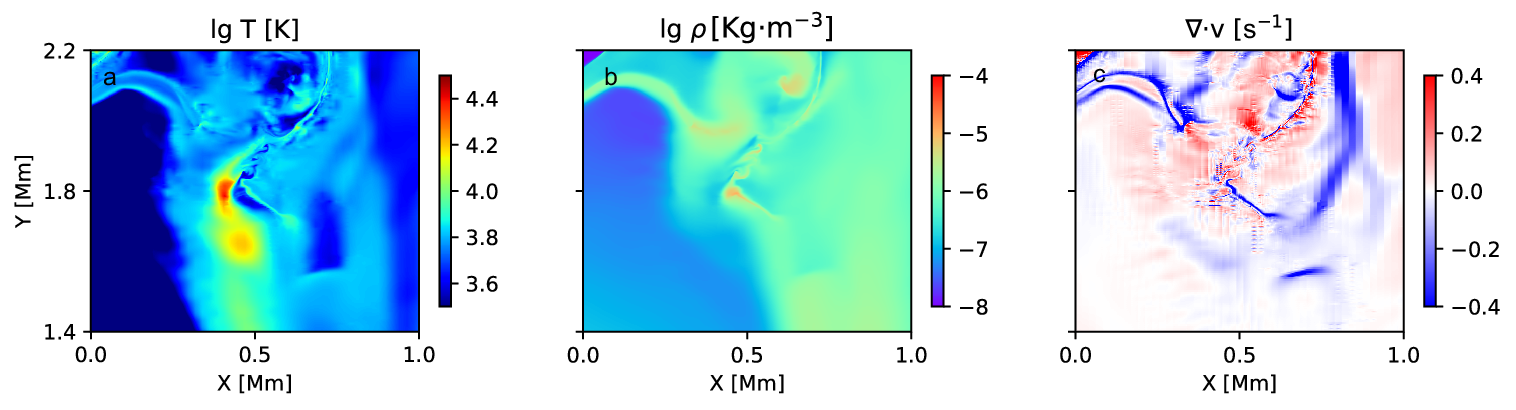}
        \caption{Zoomed-in view of turbulent reconnection outflow zone at t = 423s. Panel a shows the temperature distribution; panel b shows the density distribution; panel c shows the distribution of the velocity divergence.}
        \label{FigGam}%
\end{figure*}

           \begin{figure}
   \centering
   \includegraphics[width=8cm]{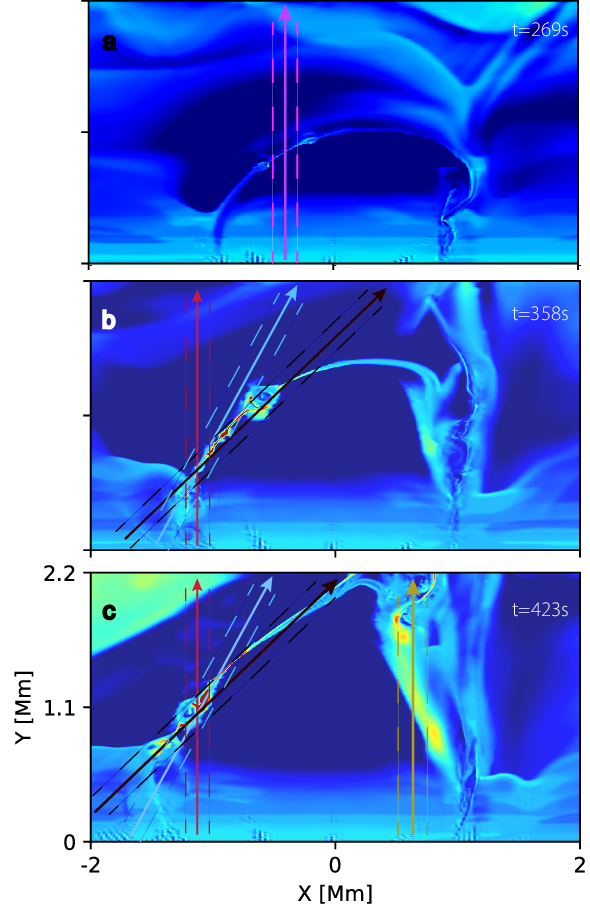}
      \caption{Diagram illustrating regions and different lines of sight for calculating the averaged spectral line profiles shown in the following five figures. The color contours represent the temperature distributions.}
         \label{FigVibStab}
   \end{figure}

             \begin{figure*}
        \centering
        \includegraphics[width=18cm]{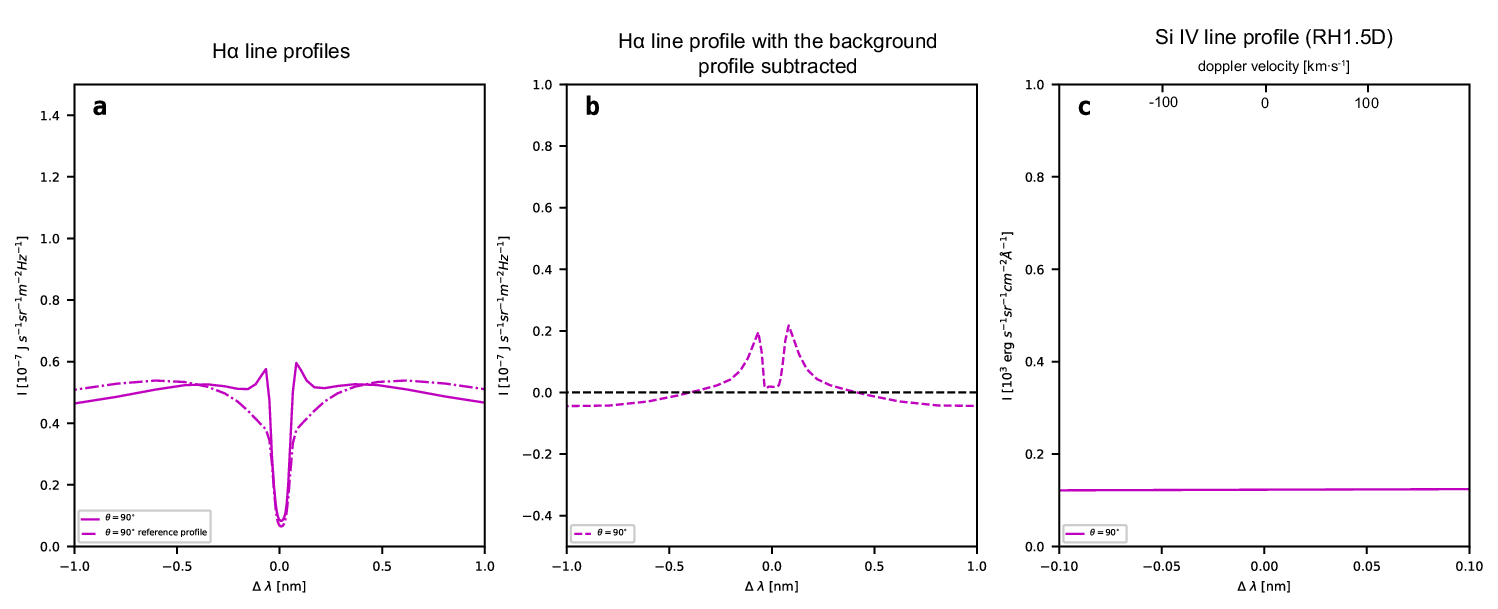}
        \caption{Synthesized spectral line profiles along the pink arrow shown in Fig 6(a) at t=269 s by using the RH1.5 code. The H$\alpha$ spectral line profile (pink solid line) and the background profile (pink dot-dashed line) are shown in (a), the H$\alpha$ spectral line profile found by subtracting the background line profile (pink dashed line) is shown in (b), and the Si IV spectral line profile (pink solid line) is shown in (c). }
        \label{FigGam}%
\end{figure*}

   \begin{figure*}
        \centering
        \includegraphics[width=18cm]{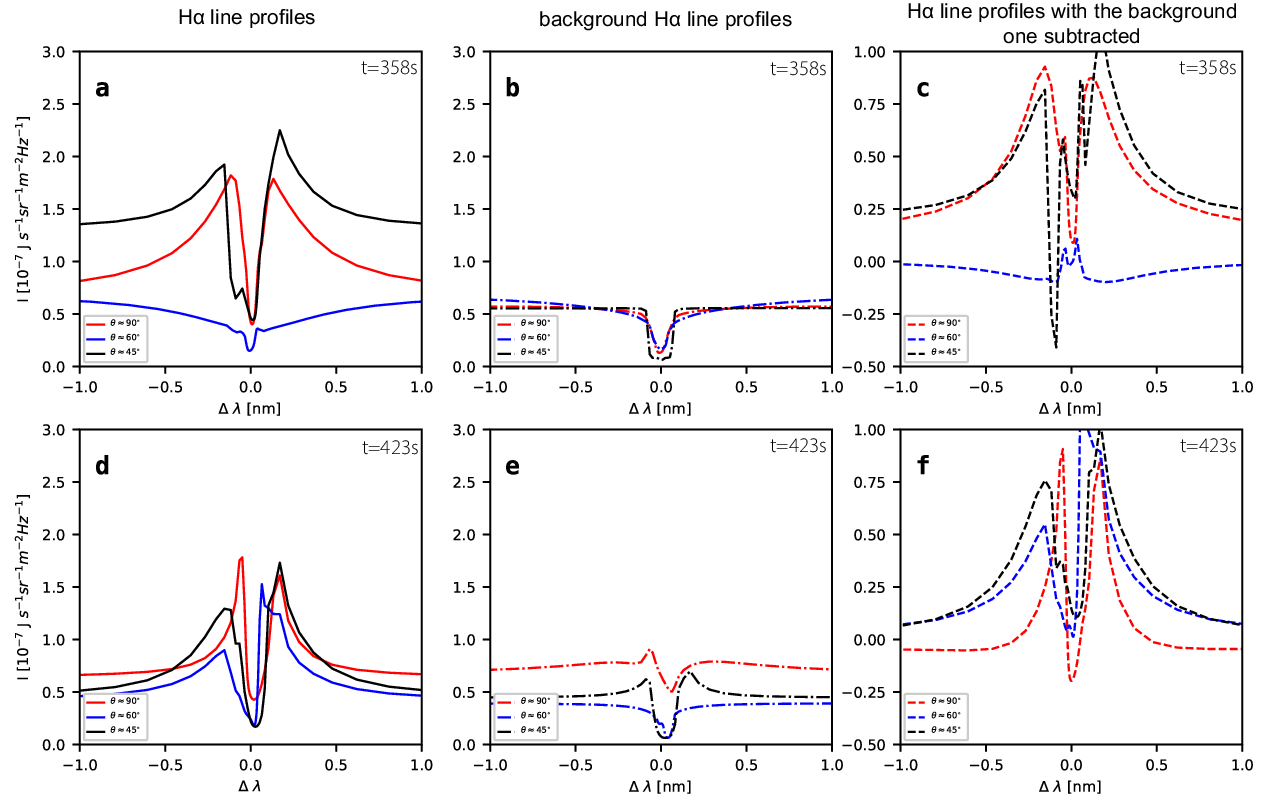}
        \caption{Synthetic H$\alpha$ spectral lines at two different times. Panels a and d show the synthetic H$\alpha$ spectral line profiles along the three different lines of sight (see Fig. 3; the angles are those between the arrows and the horizontal direction.), panels c and f show the background profiles, and panels c and f show the profiles after the ambient background profile is subtracted.}
        \label{FigGam}%
\end{figure*}

   \begin{figure*}
        \centering
        \includegraphics[width=16cm]{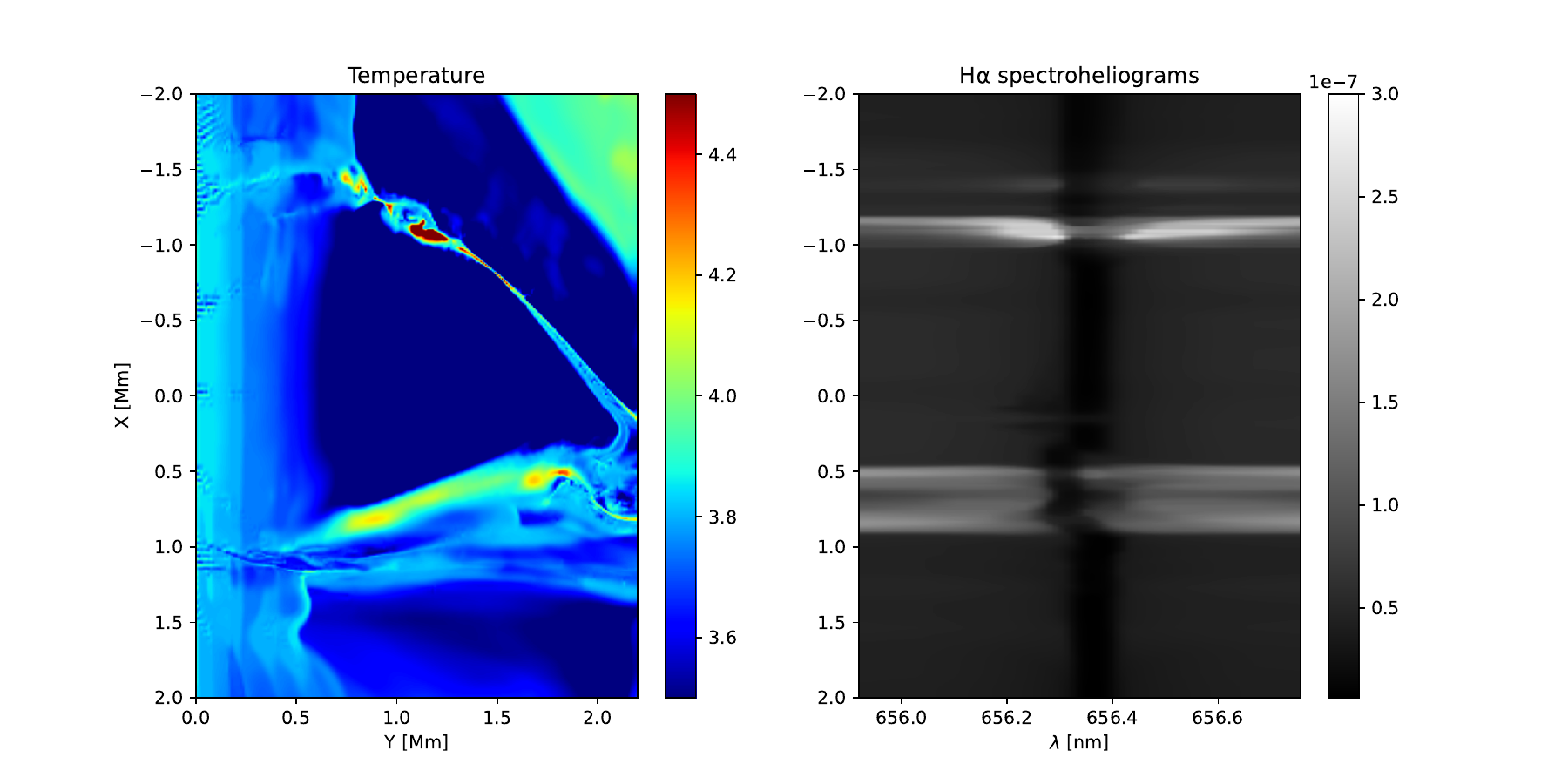}
        \caption{Temperature distribution and synthetic H$\alpha$ Spectrogram at t=423s. Panel (a) displays the temperature distribution, rotated 90 degrees for better correlation with the location of EBs, with the horizontal axis representing height (y-direction) and the vertical axis representing length (x-direction). Panel (b) shows the spectrogram synthesized to mimic Ellerman's 1917 observations, with the horizontal axis indicating wavelength and the vertical axis indicating length.}
        \label{FigGam}%
\end{figure*}

\begin{figure*}
        \centering
        \includegraphics[width=18cm]{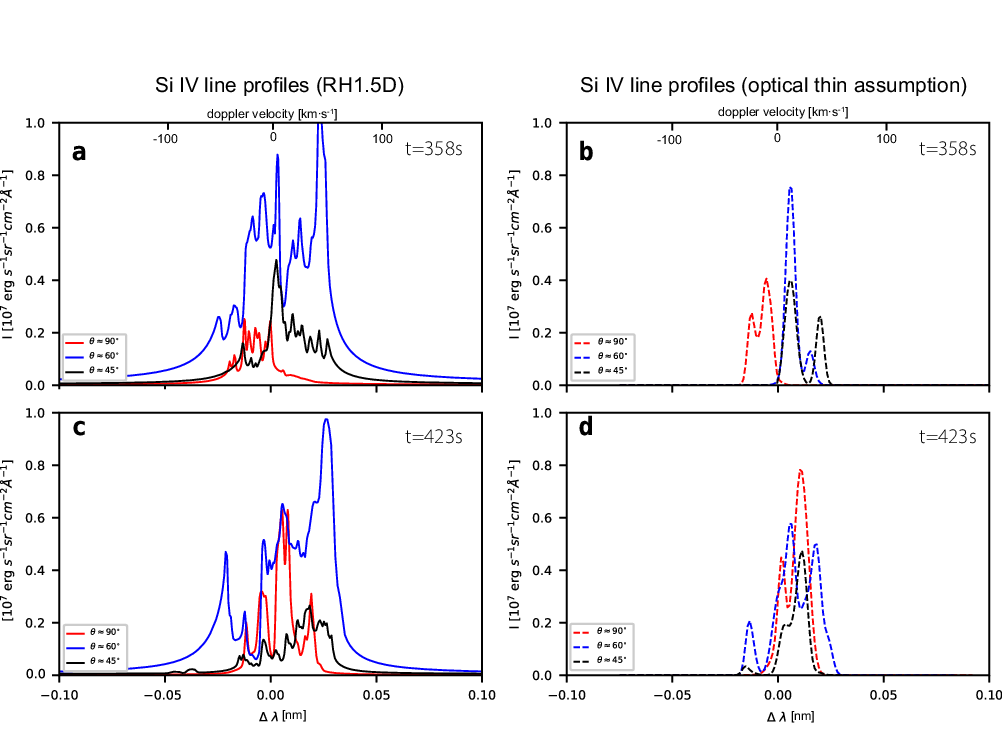}
        \caption{ Synthetic Si IV 139.4 nm line profiles at two different times along the three different lines of sight (see Fig. 3; the angles are those between the arrows and the horizontal direction). The results derived by using the RH1.5D code are shown in (a) and (c), and the results obtained by using the optically thin approximation are presented in (b) and (d).  }
        \label{FigGam}%
\end{figure*}

\section{Results}

%
\subsection{The evolution process of the emerging current sheet}

The preset arch magnetic field gradually rises and reconnects with the opposite uniform magnetic field inserted in the background. Fig. 3 displays the distributions of the plasma's temperature, density, and current density and the joint probability density function (PDF) of temperature and density at four distinct times. The whole reconnection process can be roughly divided into two stages.\\

In the first stage (t$\leq$269 s),  the arching magnetic field keeps moving upwards, and the high density plasma of the photosphere is also brought high up into the reconnection area. The plasma density in the current sheet region is comparable to that of a photosphere, and the number density of  hydrogen can reach up to 10$^{23}$ m$^{-3}$, with a plasma $\beta$ much larger than 1. It is also clear from panel h in Fig. 3 that not much high-temperature plasma is present at this moment. The average temperature in the current sheet region is only around 6000 K. Though the plasmoid instability appears later, the maximum temperature is still smaller than 10,000 K during this stage. Therefore, magnetic reconnection only results in the formation of cool EBs and no UV bursts appear in first stage.\\

In the second stage (t$\textgreater$269 s), as the simulation continues, the elongated current sheet is lifted higher, and the density gradually decreases under the effect of gravity. The plasmoid instability results in the nonuniform distributions of the plasma density, and the hot plasmas with a temperature > 20,000 K start to appear in the turbulent reconnection regions, where the plasma density is relatively lower.  As the plasmoid instability developes further, the plasmoids collide and coalesce with the nearby one, the reconnection region becomes more turbulent. The top of the current sheet is lifted to a higher altitude at the same time, and more low-density regions with hot plasmas appear. Panel p of Fig. 3 shows that the current sheet region has a large proportion of high-temperature components (>20,000K), with the plasma desnity in a range from $\sim$ 10$^{19}$-10$^{21}$ m$^{-3}$. The temperature inside the current sheet is very nonuniform, the maximum temperature can reach up to 100, 000 K, and the minimum temperature is only about 3,000 K. Therefore, we expect that EBs and UV bursts both appear during this second stage.\\

In this simulation work, the mixing of plasmas with different temperatures and densities causes the formation of EBs and UV bursts simultaneously. The multi-thermal structures manily concentrate inside the plasmoids.  Additionally, the interaction between the reconnection outflows with background plasmas and magnetic fields leads to a more complex and chaotic motion as well as an inhomogeneous temperature distribution on the right hand side of the main current sheet.\\

Figure 4 shows the specific structure of a plasmoid near the bottom of the reconnection region at t=423s, including the temperature distribution, density distribution, and distribution of the velocity divergence. We find that the region with a temperature higher than 20,000 K can be located at a low altitude around y=0.7 Mm, the high-temperature and low-temperature plasmas appear in the same plasmoid, which implies that the hot UV burst and much cooler EB can appear at a similar height.  The density of the high-temperature region is relatively low; the number density can be as low as $10^{19}$ m$^{-3}$. The velocity divergence presented in Fig. 4 reaches high values and becomes very turbulent in the reconnection region, which indicates that the local compression heating becomes important. The previous numerical results in \cite{Ni2022} highlighted that the local compression heating triggered by plasmoid instability is the dominant mechanism to heat plasmas in the UV bursts.   \\

Figure 5 shows the distributions of the temperature, density, and velocity divergence in the upper outflow region at t = 423 s. We find that a turbulent structure prevails here, with nonuniform temperature and density distributions, where the highest temperature can reach over 20,000 K and the lowest temperature is only 3,000 K. This chaotic temperature structure also allows the simultaneous appearance of UV bursts and EBs. \\

We should note that  the emergence of new arch-shaped magnetic fields from the bottom boundry propels the high-density plasma in the photosphere upwards. This movement causes the original lower chromospheric and photospheric layers shifting to higher altitudes, the optical depth of 6.5 $\times$10$^{-5}$  can even appear at about 2Mm in some regions, as shown in the first column in Fig. 3. This study employs the OPAL database to calculate the Rosseland mean opacity and optical depth across the simulation area. The choice of  6.5 $\times$10$^{-5}$  as a boundary value is based on its corresponding height of around 1.1Mm in the C7 model, where it is considered as the location at the bottom of middle chromosphere. It is acknowledged that this is not a strict demarcation but rather an indicator of the overall upward shift of the simulation area, as inferred from the relative changes in the Rosseland mean optical depth. In Figures 3(a), (e), (i), and (m), this boundary is depicted as a white dashed-dotted line. Initially positioned at y=1.1 Mm, the boundary later moves to a higher altitude. The entire reconnection area is essentially always below this boundary, indicating that the UV bursts occur in the lower chromosphere. In the next section,  we also plot the locations where the H$\alpha$ line core and wing have an optical depth of 1, further supporting our viewpoint.



\begin{figure*}
        \centering
        \includegraphics[width=18cm]{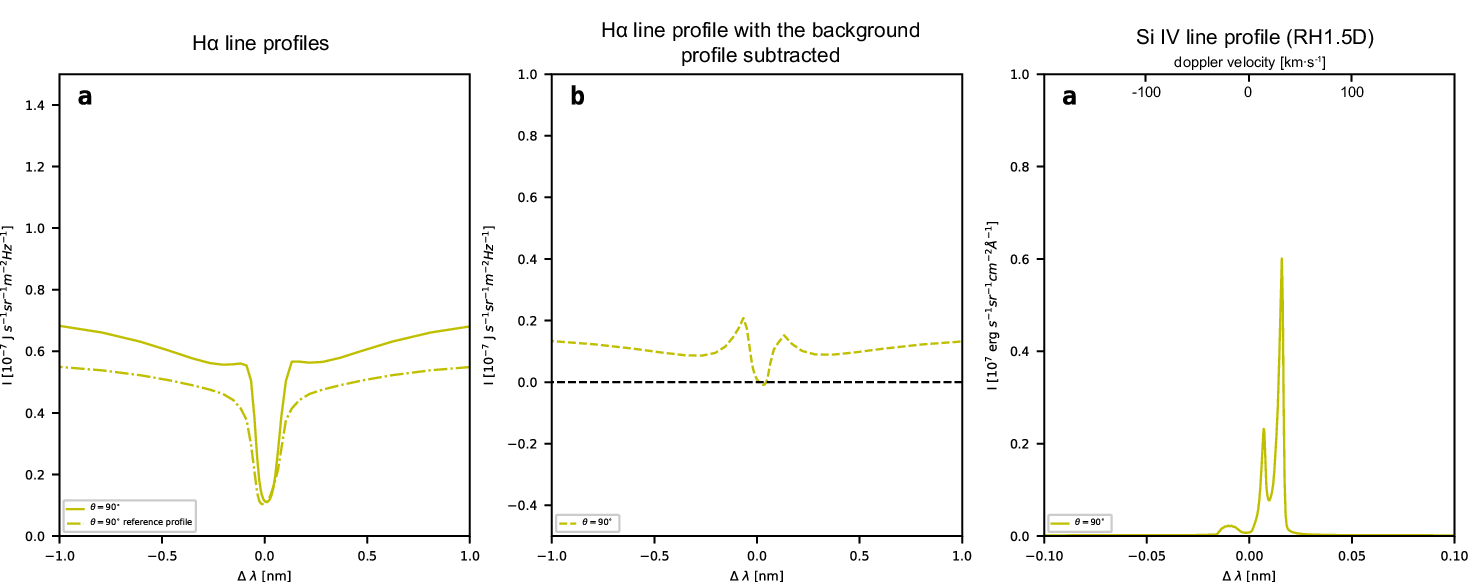}
        \caption{Synthetic spectral line profiles in the outflow region at t=423s along the yellow arrow in Fig. 6. Panel a shows the H$\alpha$ spectral line profile and the reference profile, b shows the result of subtracting the background profile from a, and c shows the profile of Si IV 193.4 nm line. These results are derived by using the RH1.5D code.}
        \label{FigGam}%
\end{figure*}

\begin{figure*}
        \centering
        \includegraphics[width=18cm]{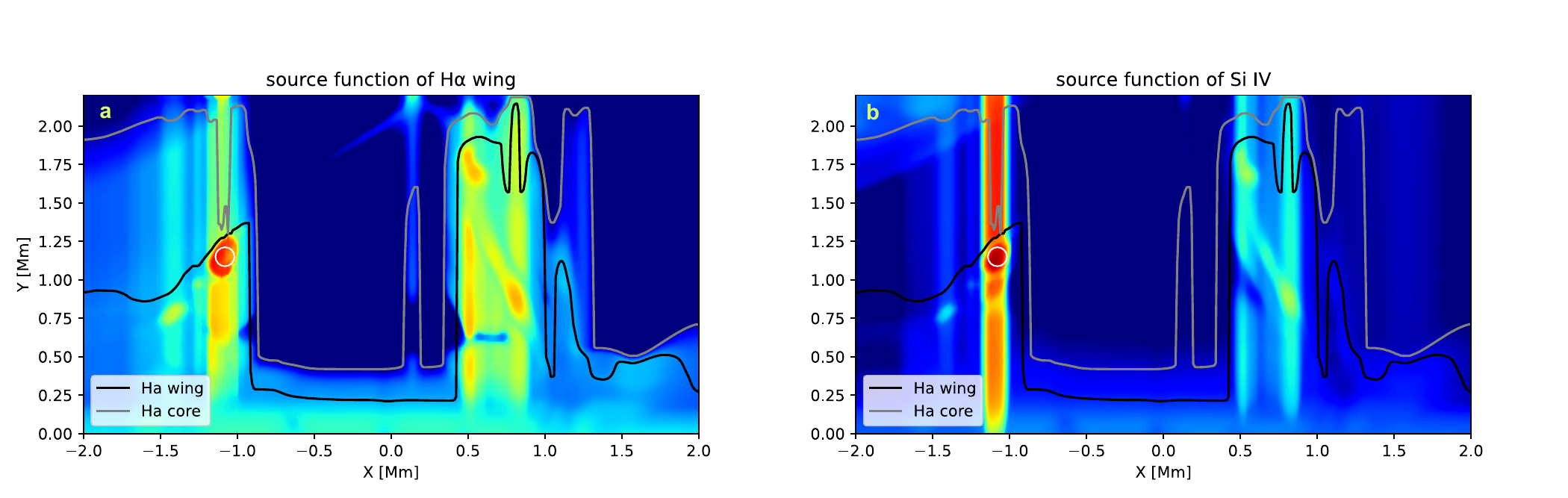}
        \caption{Distribution of source functions in simulation area at t=423s. Panel (a) presents the distribution of source functions for the H$\alpha$ wings, while panel (b) shows the source function distribution for Si IV at 139.4 nm. The gray line indicates the location where the optical depth is 1 in the H$\alpha$ line core, the black line represents the location where the optical depth of the wings is 1, and the white circles serve as reference markers. Lower values are represented in blue, transitioning through green to red for higher values. }
        \label{FigGam}%
\end{figure*}

\subsection{The synthesized H$\alpha$ and Si IV spectral lines.}

We synthesized H$\alpha$ and Si IV spectral line profiles based on our MHD simulation results by using the RH1.5D code \citep{Pereira2015}, a radiative transfer solver that is developed from a branch of the RH \citep{Uitenbroek2001} code and has better convenience, in order to more accurately diagnose EBs and UV bursts in the simulations and to better compare with the observations. In our previous related works, we synthesized the Si IV 139.4 nm spectral line profiles using the optically thin approximation \citep{Ni2021, Ni2022}; we employed the same method to calculate the Si IV spectral lines as a reference in this work. We synthesized the spectral line profiles along several different lines of sight at three different times in order to better understand the radiation properties of the reconnection region. Fig. 6 illustrates the regions and the different lines of sight for calculating the averaged spectral line profiles shown in the following four figures (Figs. 7, 8, 10, and 11). The arrows with different colors represent the different lines of sight. The angle between the pink arrow in Fig. 6(a) and the horizontal direction (x direction) is $\theta$ = 90$^\circ$, the same as the angle between the red arrow in Fig. 6(b) and the horizontal direction, and the angle between the yellow arrow in Fig. 6(c) and the horizontal direction. The angle between the blue arrow and the horizontal direction is $\theta$ = 60$^\circ$, and the angle between the black arrow and the horizontal direction is $\theta$ = 45$^\circ$. The regions between the two dashed lines on each side of the arrows are chosen to calculate the averaged spectral line profiles shown in Figs. 7, 8, 10, and 11. \\

Figure 7 presents the synthesized spectral line profiles along the pink arrow shown in Fig. 6(a) at t=269 s using the RH1.5 code. The solid pink line in Fig. 7(a) is the synthesized H$\alpha$ spectral line  profile, and the dashed pink line in Fig. 7(b) is the one obtained by subtracting the background line profile. We can find that the emission intensity and the line profile structures in Figs. 7(a) and 7(b) are very close to the observational results of a typical EB. The pink solid line in Fig. 7(c) is the synthesized Si IV spectral line profile. Such a flat line and weak emission intensity indicate that there are no UV emissions from the reconnection region at t=269 s. Therefore, only the EB-like event appears in the reconnection region before t=269 s.\\

The synthesized H$\alpha$ spectral line profiles through the reconnection region at t=358 s and t=423 s are displayed in Fig. 8. The three solid lines with different colors in Figs. 8(a) and 8(d) represent the line profiles from three different lines of sight as shown in Figs. 6(b) and 6(c), the three dashed-dotted lines in Figs. 8(b) and 8(e) are the corresponding background profiles, and the three dashed lines in Figs. 8(c) and 8(f) are the corresponding ones obtained by subtracting the background line profiles. We find that the emission intensities on the wings of all the synthetic H$\alpha$ spectral line profiles through the reconnection region are much stronger than those through background environments. Though the values of the emission intensity and the widths of the blueshift and redshift are not exactly the same when the lines of sight are different, all the synthetic H$\alpha$ spectral line profiles with the background profiles subtracted have the mustache shaped structure and a typical intensity of 10$^{-8}$ J s$^{-1}$ sr$^{-1}$ m$^{-2}$ Hz$^{-1}$, which match well with the observed characteristics of an EB \citep[e.g.,][]{Pariat2007}. Therefore, the cold parts of the reconnection region still show the EB-like features, even though the hot plasmas appear during this period.   \\ 

In Fig. 9, we present the temperature distribution of the simulation area at t=423s (Fig. 9a), along with the H$\alpha$ spectrogram computed for a line of sight perpendicular to the solar surface (Fig. 9b). The horizontal and vertical axes of Fig. 9a represent the height and length of the simulated region, respectively, while in Fig. 9b, the horizontal axis covers the wavelength range around H$\alpha$ 656.3 nm, and the vertical axis represents the length of the simulated region in the x direction. Notably, Fig. 9b presents a similar result to the one based on the observations of 1917 \citep{Ellerman}, showing a clear enhancement in the wings of the H$\alpha$ line. This corresponds to to the reconnection region with multiple islands and the outflow region at the bottom in Fig. 9a, with no significant enhancement or only weak enhancement at the line core. This also suggests that the reconnection region and the outflow region both exhibit clear observational characteristics of EBs.\\

Figure 10 displays the line profiles of Si IV 139.4 nm at t=358s and t=423 s, the results derived by using the RH1.5D code are presented in Figs. 10(a) and 10(c) in the left panel, Figs. 10(b) and 10(d) in the right panel show the results obtained by using the optically thin approximation. We can find that the two different methods cause obvious differences on the synthetic line profiles. When the radiative transfer code RH1.5D is applied, the widths of these line profiles are wider and they have more spikes, as shown in Figs. 10(a) and 10(c). The low solar chromosphere is not completely optically thin. Therefore, the optically thin approximation method will cause the synthetic Si IV spectral line as shown in Figs. 10(b) and 10(d) to more or less deviate from the realistic ones on some aspects. However, with the general properties of the synthetic line profiles found using two different methods, such as the intensity, the redshift and blueshift structures are similar, and they all have multi-peaked structures. As shown in Fig. 10, the magnitude of the emission intensity is ~ 10$^{6}$ erg s$^{-1}$ sr$^{-1}$ cm$^{-2}$ $\AA^{-1}$ , and the width of the line profile can reach from -100 to 100 km s$^{-1}$. These characteristics of the synthetic Si IV spectral lines are very similar to the observational results of the typical UV burst \citep[e.g.,][]{Young}. It should be noted that some UV bursts exhibit Si IV line widths reaching from -200 to 200  km s$^{-1}$, a velocity that is challenging to achieve in the low chromospheric environment of our simulation due to its higher density. However, the line width from -100 to 100 km s$^{-1}$ still falls within the range of UV bursts or similar events. The line width (~0.1 nm) shown in Figs. 10(a) and 10(c) is close to that (~0.11nm) presented in Fig.7 in \cite{Hansteen2019}; the Ni II absorption line was successfully synthesized in their paper. The previous observational results \citep{Peter} showed that the Ni II absorption line appears at around -90 km s$^{-1}$. Therefore, the width from -100 to 100 km s$^{-1}$ may still have a chance to find the Ni II absorption line. As shown in Figs. 3(i) and 3(m), plenty of dense plasmas with temperatures below 6500 K are located above the hot UV emission, which also support the possible generation of Ni II absorption. In the future, we hope that we can have a chance to obtain the Ni II atomic table in order to check this viewpoint.\\

We note that the reconnection outflows continue to collide with the background plasmas and magnetic fields, which then result in the formation of a chaotic region on the right side of the main current sheet (see Fig. 5). Many turbulent structures with nonuniform density and temperature distributions appear in this region, the temperature of the hot plasmas can reach above 20,000 K. We calculated the spectral line profiles of H$\alpha$ and Si IV 139.4 nm through this region by using the RH1.5D code, the line of sight is along the yellow arrow shown in Fig. 6(c), and the results are presented in Fig. 11. The synthetic H$\alpha$ spectral line profiles shown in Figs. 11(a) and 11(b) match very well with the observed one for a typical EB. From Fig. 9(b), we can also draw a parallel conclusion.  Fig. 11(c) indicates that the hot plasmas in this region also  bring about the strong Si IV line emissions. These results demonstrate that the UV bursts and EBs can also simultaneously appear in this turbulent region caused by the reconnection outflows. \\

In this simulation, we focus on the formation heights of EBs and UV bursts. Fig.12 respectively presents the distributions of the source functions for the H$\alpha$ wings and Si IV at t = 423 s, calculated using RH1.5D from a line of sight perpendicular to the solar surface. Additionally, the black and gray lines in Fig. 12 correspond to the locations where the optical depth equaled 1 for the H$\alpha$ wing and core, respectively. EBs are characterized by a lack of significant response in the line core and enhanced emission in the wings. The distribution of the source functions for the H$\alpha$ wings, influenced by radiation from other regions and calculated under nonlocal thermodynamic equilibrium (NLTE) conditions, suggests that the areas with notably strong contributions along the line of sight are crucial. As shown in Fig. 12, most of the H$\alpha$ wing and Si IV emission sources are located below the black line where $\tau =1$ for the line wing of H$\alpha$, the maximum value of the Si IV source function is basically located inside the small white circle and the maximum value of H$\alpha$ wing source function is surrounding the white circle, but they are very close. Such a result further supports our perspective that the UV emissions and H$\alpha$ wing emissions can be generated in the same low atmospheric layer, and they can be mixed together in the turbulent reconnection region.\\


\section{Conclusions}
In this work, we used high-resolution MHD numerical simulations to explore the reconnection process of the emerging magnetic field with the background field and analyze the formation mechanisms of EBs and UV bursts as well as the relationship between them. Compared to the previous work \citep{Ni2021}, the time-dependent ionization degree was included in order to obtain the more realistic diffusivities, and the more realistic radiative cooling model was applied. In addition, we used the RH1.5D code to synthesize the H$\alpha$ and Si IV spectral line profiles or spectrograms through the simulated region, and we compared the results with observations. We also provided the source functions for H$\alpha$ wing radiation and Si IV radiation, as well as the positions of the unit optical depth for both the H$\alpha$ line core and wings. This information aids in the determining of the formation regions and altitudes of EBs and UV bursts in our simulation. The main conclusions are as follows. \\

   \begin{enumerate}
      \item During the flux emergency process, the reconnection current sheet between the emerged and background magnetic fields is gradually reaching a higher altitude. Firstly, the EB is formed inside the reconnection region, and the hot UV burst appears a few tens  of seconds later, which is consistent with the recent observations \citep{Ada2020}. The current sheet is always located inside the environment with dense plasmas and below the boundary with a Rosseland mean optical depth value~of 6.5$\times$ 10$^{-5}$, and the primary contributions to the H$\alpha$ wing radiation and Si IV radiation predominantly come from the region below the height where the optical depth of the H$\alpha$ wing equals 1. Therefore, we predicate that EBs and UV bursts in our simulations are always located below the middle chromosphere.

      \item After the plasmoid instability appears, the turbulent reconnection process makes the hot plasmas with low density and the cool plasmas with high density to alternately appear in different locations inside the main current sheet. The hot plasmas with a temperature of ~100,000 K and the cool plasmas with a temperature below 10,000 K can simultaneously appear even in one plasmoid.  Passing through the turbulent reconnection region, the synthetic Si IV spectral line profile shows multiple peaks, and the width can reach from -100 to 100  km s$^{-1}$. The synthetic H$\alpha$ spectral line profile with strong wing emission and absorption at the line center also match well with a typical EB. These results confirm that UV bursts and EBs can be mixed in the turbulent reconnection region at a similar height.

      \item The region where the reconnection outflow interacts with the background magnetic fields and plasmas is also turbulent and filled with plasmas with different densities and temperatures. The synthetic Si IV and H$\alpha$ spectral line profiles through this region indicate that UV bursts and EBs can also be mixed up in this region.    \end{enumerate}

These results further prove that the reconnection model we have proposed in this and our previous work \citep{Ni2021} can well explain the UV bursts connecting with EBs. The model proposed by us is  different from that in \cite{Hansteen2019}, the UV emission in our simulation can extend downward to a lower height at around 0.5 Mm. More importantly,  they concluded that EBs and UV bursts correspond to reconnection at different atmospheric layers in \cite{Hansteen2019} and \cite{Ada2020}. The coexisting EBs and UV bursts would be part of the same reconnection system, but happening far apart vertically. However, the whole curved current sheet in our model is basically always below the middle chromosphere (the white dashed-dotted line). As shown in Fig. 3, the hot plasmas above 20,000 K and the much colder plasmas below 10,000 K are mixed together, and they alternately appear in space in the reconnection region as shown in Fig. 3 and Fig. 4. The hot sources corresponding to UV emissions can appear at about the same height as the colder sources corresponding to EBs, and the hot sources can also appear at an even lower height than the colder sources. Such a scenario is obviously different from the results and conclusions shown in \cite{Hansteen2019} and \cite{Ada2020}. On the other hand, our simulations clearly present the turbulent reconnection mediated with multiple plasmoids; the hot plasmas and much colder plasmas can even appear in the same plasmoid (Fig.4), which was not shown in \cite{Hansteen2019}.
\\

The multi-thermal structures in the plasmoids along the corona jet have been identified in multi-wave lengths high resolution observations \citep{Zhang2019}. Therefore, we expect that the similar phenomenon can also occur in the smaller scale events such as UV bursts and EBs, which requires the future high-resolution solar telescopes with multiple wavelengths to prove this point.\\


We should mention that the hydrogen gradually becomes closer to the LTE when the plasma is heated to a high temperature; the \cite{Carlsson2012} model then overestimates the radiative cooling effect of \cite{Hong2022}. Thus, more plasmas  than those shown in our simulations are expected to be heated up to the higher temperatures in the turbulent reconnection region.\\

The Si IV line profiles calculated with RH1.5D can reach from -100 to 100 km s$^{-1}$ in this work, which is close to the synthesized result in \cite{Hansteen2019}, and they are wider than those in \cite{Hong2022}. After we obtain the Ni II atomic table, it is necessary to check if the Ni II absorption line can be derived in the next work. In the future, we will perform the 3D high-resolution MHD simulations with the more realistic radiative cooling model to study this topic. The convection zone with the radiative transfer process will be included to result in the more realistic flux emergence process. We will also synthesize the emission intensity images and spectral line profiles with different wavelengths to better compare with observations.

\begin{acknowledgements}
We would like to thank the referee for his/her insightful and constructive comments for improving our paper. We thank Jie Hong for providing the Si IV model atom and for discussions about the radiative cooling model. We thank Feng Chen for the discussions about the radiative cooling model. This research was supported by the National Key R$\&$D Program of China Nos. 2022YFF0503003 (2022YFF0503000) and 2022YFF0503804 (2022YFF0503800); the NSFC Grants 11973083 and 11933009; the Strategic Priority Research Program of CAS with grants XDA17040507; the outstanding member of the Youth Innovation Promotion Association CAS (No. Y2021024 ); the Applied Basic Research of Yunnan Province in China Grant 2018FB009; the Yunling Talent Project for the Youth; the project of the Group for Innovation of Yunnan Province grant 2018HC023; the Yunling Scholar Project of the Yunnan Province and the Yunnan Province Scientist Workshop of Solar Physics; Yunnan Key Laboratory of Solar Physics and Space Science under the number 202205AG070009; The numerical calculations and data analysis have been done on Hefei advanced computing center and on the Computational Solar Physics Laboratory of Yunnan Observatories.
\end{acknowledgements}

%
%

\end{document}